\documentclass[12pt,twoside]{article}
\usepackage{fleqn,espcrc1}
\usepackage{latexsym} 
\usepackage{amssymb}  
\usepackage{amsfonts} 
\usepackage{multirow}
\usepackage{epsf}
\usepackage{graphicx}
\def\y0{y^{(0)}}

\newcommand \beq{\begin{eqnarray}}
\newcommand \eeq{\end{eqnarray}}

\newcommand{\mnote}[1]{\marginpar{\tiny {}}}   

\begin{document}

\title{\bf Chemical Equilibration and the Hadron-QGP Phase Transition
}
\author{
   Peter Braun-Munzinger\\
   Gesellschaft f{\"u}r Schwerionenforschung\\
   64220 Darmstadt, Germany\\
   }
\maketitle

\begin{abstract}

    \noindent We discuss recent experimental results on hadron multiplicities
in     ultra-relativistic nuclear collisions. The data for central
collisions are in  quantitative agreement with predictions of a
thermal model assuming full chemical equilibration. It is argued that
this provides strong, albeit indirect, evidence for the formation of a
partonic phase in the collision prior to hadron production.

\end{abstract}


\section{Introduction} 
\noindent Experiments  with ultra-relativistic nuclei are performed to
produce and study the quark-gluon plasma. This new state of
matter is predicted to exist at high temperatures and/or high baryon
densities. Numerical solutions of QCD using lattice
techniques imply that the critical temperature (at zero baryon
density) is about 170 MeV \cite{lattice}. Comprehensive surveys of the
various experimental approaches on how to produce such matter in
nucleus-nucleus collisions have been given recently
\cite{pbmpanic,jsinpc,pbmqm97,bass}.  Here we focus exclusively on
hadron production, its interpretation in terms of a thermal model,
and the resulting consequences for the quark-hadron phase
transition.

\section{Thermal Model, Strangeness Enhancement and Equilibration}
\noindent 
The statistical model used here is presented in detail in
\cite{therm3}. Like its predecessors  presented
in \cite{therm2,therm1} it is based on the use of a grand canonical ensemble
to describe the partition function and hence the density of the
particles of  species \(i\) in an equilibrated fireball:

\begin{equation}
n_i= \frac{g_i}{2 \pi^2} \int_0^\infty \frac{p^2 \, {\rm
d}p}{e^{(E_i(p)-\mu_i)/T} \pm 1}
\label{grundgl}
\end{equation}

\noindent with \(n_i\) = particle density, \(g_i\) = spin degeneracy,
\(\hbar\) = c = 1, \(p\) = momentum, \(E\) = total energy and chemical
potential \(\mu_i = \mu_B B_i+\mu_S S_i+\mu_{I_3} I^3_i\). The
quantities \( B_i\), \( S_i\) and \( I^3_i\) are the baryon, strangeness
and three-component of the isospin quantum numbers of the particle of
species \(i\). The temperature T and the baryochemical potential
\(\mu_B\) are the two independent parameters of the model, while the
volume of the fireball V, the strangeness chemical potential \(\mu_S\),
and the isospin chemical potential \(\mu_{I_3}\) are fixed by the three
conservation laws for baryon number, strangeness, and charge.
Interactions among hadrons are taken into account via an excluded
volume correction. For details see \cite{therm3}.

\begin{figure}[thb]

\vspace{-1cm}

\epsfxsize=15cm
\begin{center}
\hspace*{0in}
\epsffile{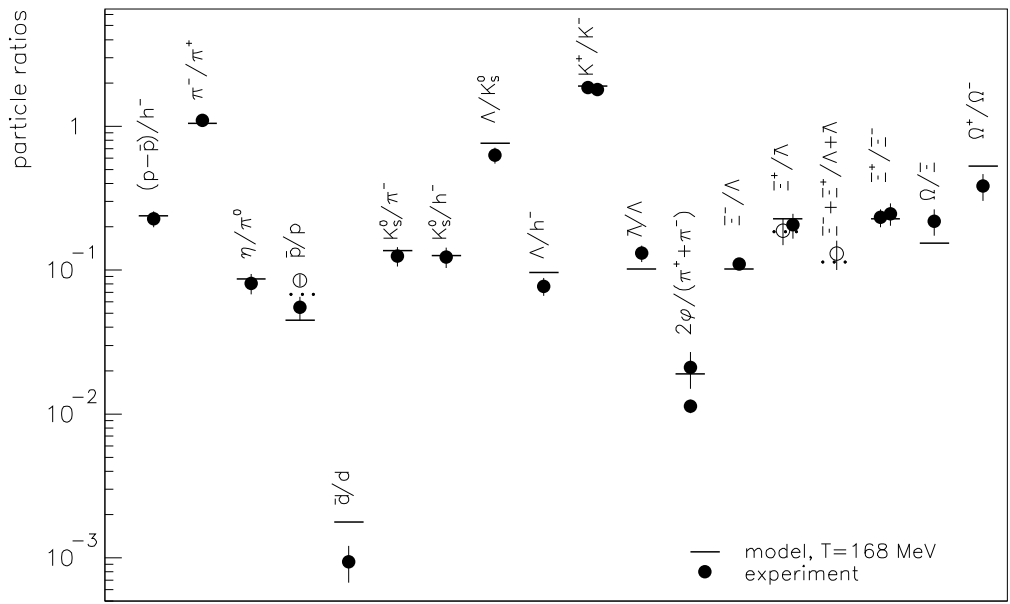}
\end{center}

\vspace{-1.5cm}

\caption{
Comparison of measured particle ratios with predictions of the thermal
model. For details see text and \cite{therm3}.
} 
\label{fig:ratios}
\end{figure}

The aim of this approach is to determine whether or not the observed
hadron yields can be described in a model assuming complete chemical
equilibration. Note that temperature and baryon chemical potential are
fixed by the pion/baryon and the anti-nucleon/nucleon ratios, and
hence no information about the yields of strange particles is used to
determine the two parameters of the model. 
The production yields of strange hadrons are, however,  significantly
increased in ultra-relativistic nuclear collisions compared to what is
expected from a superposition of nucleon-nucleon collisions. This has
been observed by several experiments both at the AGS and at the
SPS \cite{qm99}. Nevertheless, all hadron yields including those for
multi-strange baryons, where enhancement factors of more than 1 order
of magnitude are observed,  can be described consistently
\cite{therm3} if one assumes a 
fireball with temperature T=168 Mev and baryon chemical potential
$\mu_b$ = 266 MeV. This is demonstrated in Fig. ~\ref{fig:ratios}. On
the other hand, the observed enhancement, especialy for 
multistrange hadrons, cannot currently  be understood within any of the
hadronic event generators \cite{qm99}

How chemical equilibration can be reached in a purely hadronic
collision is not clear in view of the small production cross section
for strange and especially multi-strange hadrons. In fact, system
lifetimes of the order of 50 fm/c or more are needed for a hot
hadronic system to reach full chemical equilibration
\cite{life}. Such lifetimes are at variance with lifetime values
established from interferometry analyses, where upper limits of about
10 fm/c are deduced \cite{appels}.

\begin{figure}[thb]

\vspace{-1.1cm}

\epsfxsize=11cm
\begin{center}
\hspace*{0in}
\epsffile{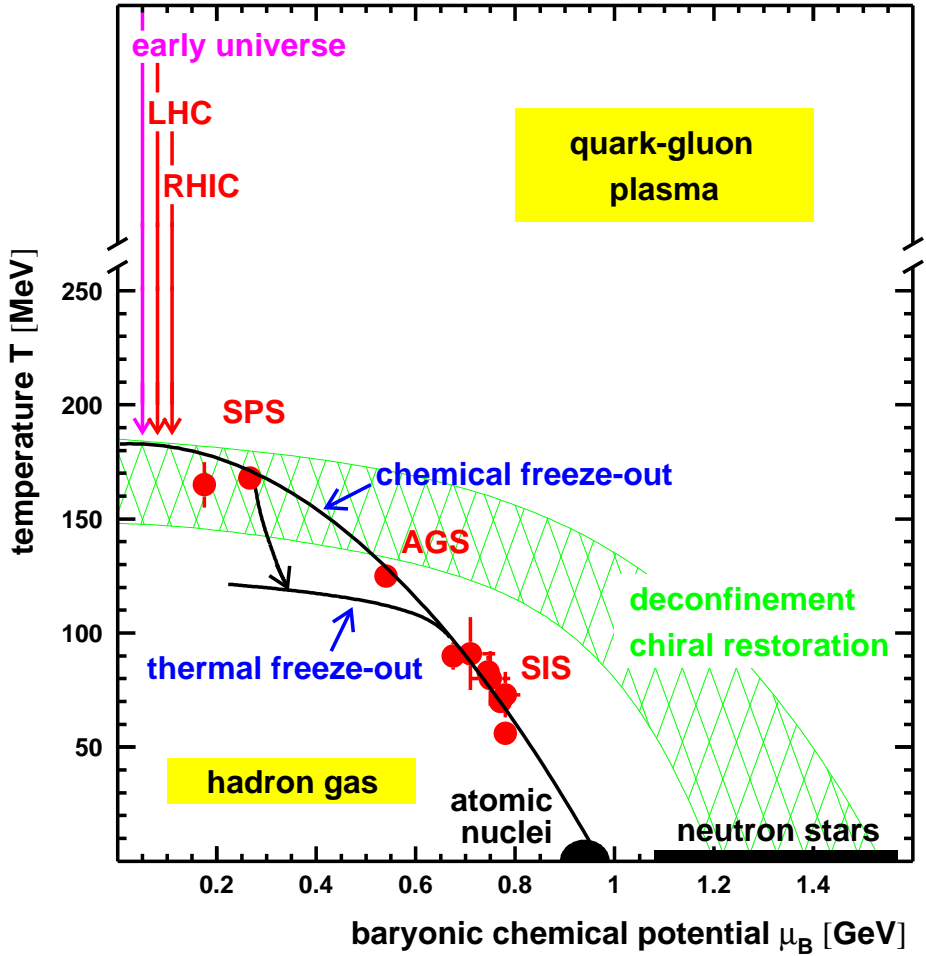}
\end{center}

\vspace{-2.0cm}

\caption{
Phase diagram of hadronic and partonic matter. The hadrochemical
freeze-out points are determined from thermal model analyses of heavy
ion collision data at SIS, AGS and SPS energy. The hatched region
indicates the current expectation for the phase boundary based on lattice QCD
calculations at $\mu_b$=0. The arrow from chemical to thermal
freeze-out for the SPS corresponds to isentropic expansion.
} 
\label{fig:phase-d}
\end{figure}

Another very interesting observation is that the chemical
potentials $\mu_b$ and temperatures T resulting from the thermal analyses of
\cite{therm3,therm2,therm1} place the systems at chemical freeze-out
very close to where we currently believe is the phase boundary between
plasma and hadrons. This is demonstrated in Fig. ~\ref{fig:phase-d}
\footnote{This is an updated version of the figure shown in
\cite{jsinpc,pbmqm97}.}  where also results from lower energy analyses
are plotted. The freeze-out trajectory (solid curve through the data
points) is just to guide the eye but follows closely the empirical
curve of \cite{freeze}.

The closeness of the freeze-out parameters (T,$\mu_b$) to the phase
boundary might be the clue to the apparent chemical equilibration in
the hadronic phase: if the system prior to reaching freeze-out was in
the partonic (plasma) phase, then strangeness production is determined
by larger partonic cross sections as well as by hadronization. Slow
cooking in the hadronic phase is then not needed to produce the
observed large abundances of strange hadrons. Early simulations of
strangeness production in the plasma and during hadronization support
this interpretation at least qualitatively \cite{knoll}. 

We would like to stress, however, that it cannot be {\it just} the
non-perturbative  hadronization mechanism which brings the hadron
abundances into ``apparent'' chemical equilibrium, as has been argued
recently \cite{stock,heinz}. The hadronization of single quarks, as
witnessed by a recent thermal analysis of hadron abundances following
e$^+$e$^-$ annihilation \cite{becattini}, leads to very drastically different
distributions: as is known for many years now, hadrons with
strangeness are significantly suppressed compared to purely thermal
expectations. This leads in \cite{becattini} to a strangeness
suppression factor $\gamma_s$ 
= 0.67, implying that $\Omega$ production is reduced by more than a factor of 3
compared to the full chemical equilibrium value reached in Pb-Pb
collisions. The role of gluons is clearly important in the partonic
state reached in heavy ion collisions.

Further strong support for a thermal interpretation of the observed hadron
yields also arises from recent results on event-by-event fluctuations
\cite{na49_qm99}. The observed distributions for the mean transverse
momentum and the kaon/pion ratio look indistinguishable from reference
distributions obtained by event mixing, implying only tiny\footnote{At
the 90 \% confidence level, nonstatistical contributions of less than
1 \% to the mean transverse momentum distributions and less than 5 \%
to the kaon/pion ratio are excluded.}
non-statistical components in these distributions. Since the kaon/pion
ratio in the current interpretation is frozen at chemical
equilibration, i.e. at the phase boundary, this result strengthens the
argument for complete chemical equilibrium.

We note also that the current thermal interpretation works most
convincingly for very central collisions. If the parameters T and
$\mu_b$ are kept constant, all thermal yields should scale linearly
with the volume and, hence, with the number of participants, implying
that all particle ratios should be independent of collision
centrality. This is indeed observed for the yields and ratios of
multi-strange baryons in the range of participant numbers larger than
100 \cite{wa97}. For pions and kaons one observes,
however, a small but significant increase of yields with the number of
participants \cite{na49_qm99}. Such an increase could imply a small
decrease of $\mu_b$ with the number of participants. However, the
observed anti-proton/proton ratios and yields of multi-strange baryons
are not in support of such an interpretation. Further work is necessary to
understand all finer aspects of hadron production.

\section{Summary and Outlook}
\noindent Hadron production results from central nucleus nucleus
collisions at ultra-relativistic energies can be quantitatively
understood by assuming that the fireball formed in the collision
freezes out chemically very near to the phase boundary between
quark-gluon plasma and hadron gas. This result cannot be explained
within purely hadronic scenarios and lends strong support to the
interpretation that the freeze-out state is reached via a system
trajectory which crosses the phase boundary from the quark-gluon
plasma side. 

This interpretation leads directly to  predictions for hadron
production at RHIC \cite{stachel_rhic} and LHC \cite{pbm_lhc} which
should be (at least for RHIC) testable soon.

\newpage

\end{document}